\documentclass[showpacs,prb,preprintnumbers,amsmath,amssymb,twocolumn]{revtex4}

\usepackage{graphicx}% Include figure files
\usepackage{dcolumn}% Align table columns on decimal point
\usepackage{bm}% bold math

\begin{document}
\title{On the extraction of paramagnon excitations from resonant inelastic X-ray scattering experiments}

\author{Jagat Lamsal and Wouter Montfrooij}
 \affiliation{Department of Physics and Astronomy, and the Missouri Research Reactor, University of Missouri, Columbia, 65211 MO.}
\begin{abstract}
{Resonant X-ray scattering experiments on high-temperature superconductors and related cuprates have revealed the presence of intense paramagnon scattering at high excitation energies, of the order of several hundred meV. The excitation energies appear to show very similar behavior across all compounds, ranging from magnetically ordered, via superconductors, to heavy fermion systems. However, we argue that this apparent behavior has been inferred from the data through model fitting which implicitly imposes such similarities. Using model fitting that is free from such restrictions, we show that the paramagnons are not nearly as well-defined as has been asserted previously, and that some paramagnons might not represent propagating excitations at all. Our work indicates that the data published previously in the literature will need to be re-analyzed with proper models.}
\end{abstract}
 \maketitle 

Recent advances in resonant X-ray scattering instrumentation\cite{rixs} have allowed for the measurement of high-energy magnetic excitations in strongly correlated electron systems, such as high-temperature superconductors. We refer to the literature\cite{rixs} for details on the technique. In these experiments, the magnetic excitations (paramagnons) show up as easily identifiable excess intensity in the scattering patterns. Typically, high-energy paramagnons show up as broad excitations peaking at finite energy transfers of a few hundred milli-electronVolt, (reasonably) well separated from elastic scattering and located on top of a sloping background. In order to extract the details of the paramagnons, such as their propagation frequencies, damping rates and intensities, the spectra are modeled by a damped harmonic oscillator function in addition to an eleastic contribution and a background contribution. These fits tend to give a good description of the experimental data.\\

However, the model that is being used to analyze resonant inelastic X-ray scattering (RIXS) data is fundamentally flawed in the sense that it imposes restrictions beforehand that are not necessarily met in the systems under study. In particular, when modeling data to a harmonic oscillator model, one should allow for the possibility that oscillations can be not only damped, but also critically damped and even overdamped\cite{wouterbook,rot}. The models used in the literature\cite{disclaimer,ybco,lasrcuo,cupa,pnictides,bis,bis2}  do not allow for the latter possibilities. In this brief report we detail the exact problem with the models in use, how the usage of such restricted models leads to fitting parameters that appear to be more accurate than they actually are, and how its usage has, in some instances, led to the identification of overdamped oscillations with being propagating paramagnons.\\

In X-ray scattering experiments one obtains information\cite{rixs} about the dynamic structure factor of the system, $S(\vec{q},\omega)$, with $\hbar \vec{q}$ the amount of momentum transferred to the system, and $E$= $\hbar \omega$ the amount of energy transferred. The measured signal not only contains the information on paramagnons contained in $S(\vec{q},\omega)$, but also unwanted scattering by the system, such as the excitation of multiple paramagnons in a single scattering event, or in separate scattering events. While it is possible to correct the data for these effects and for the scattering by sample holders and background scattering in general, it does make the analysis more intricate than a cursory inspection of the scattered data would appear to imply.\\

The dynamic structure factor is related to the imaginary part of the dynamic susceptibility $\chi"(\vec{q},\omega)$  by
\begin{equation}
\chi"(\vec{q},\omega)=(1-e^{-\beta \hbar\omega}) S(\vec{q},\omega),
\label{db}
\end{equation}
where $\beta = 1/k_B T$, with $k_B$ Boltzmann's constant and $T$ the temperature of the system. The poles of the dynamic structure factor $\chi(\vec{q},\omega)$ determine the excitations of the system. As such, in scattering experiments these poles can be masked by a temperature dependent frequency factor, and therefore, the peak positions in $S(\vec{q},\omega)$ do not necessarily correspond to the excitation energies of the paramagnons. We clarify that in the following by using the harmonic oscillator model employed in model fitting to RIXS data.\\.

When modeling the paramagnon part of the scattering in RIXS experiments, one employs\cite{ybco,lasrcuo,cupa,pnictides,bis,bis2} the following model for  $\chi"(\vec{q},\omega)$:
\begin{equation}
\chi"(\vec{q},\omega)=
\cfrac{\Gamma_{\vec{q}}}{\Gamma_{\vec{q}}^2+(\omega-\omega_{\vec{q}})^2}
-\cfrac{\Gamma_{\vec{q}}}{\Gamma_{\vec{q}}^2+(\omega+\omega_{\vec{q}})^2}.
\label{ah}
\end{equation}
In here, $\Gamma_{\vec{q}}$ and $\omega_{\vec{q}}$ are the propagation frequencies and damping rates of the paramagnons, respectively. The reader will recognize this as the Fourier transform of the solutions of the damped harmonic oscillator equation, where one obtains two solutions in the time domain, representing damped waves traveling in opposite directions. As a concrete example, picture a pendulum clock swinging in air. The air provides a damping $z=2\Gamma$, and the two possible solutions to the harmonic oscillator equation show up as the swigning of the pendulum, back and forth. When the damping is increased, we find that the oscillation frequency $\omega_{osc}$ diminishes according to $\omega_{osc}=\sqrt{f^2-z^2/4}$. Here $f$ stands for the undamped frequency. Once the damping reaches and exceeds the critical value $z= 2f$, then the oscillations stop and instead we are left with a purely damped motion, with the two solutions now described by simple exponentials with decay constants $\Gamma_{\pm} = z/2 \pm \sqrt{z^2/4-f^2}$. We would have this situation if we were to place our pendulum under water.\\

While the above is introductory undergraduate physics, it is often not realized that by employing eqn \ref{ah} one explicitely excludes the possibility that excitations can become overdamped. There are multiple ways to verify this. First, while the solutions of an overdamped harmonic oscillator (in time) are chracterized by two decay rates, eqn \ref{ah} with $\omega_{\vec{q}}$= 0 is characterized by only one damping rate. Second, substituting $\omega_{\vec{q}}$= 0 directly into eqn \ref{ah} yields a dynamic structure factor identical to zero.\\

When attempting to fit eqn \ref{ah} to scattering data, one will still obtain a reaonable fit. Since the fit is prohibited from reaching $\omega_{\vec{q}}$= 0 because this would correspond to a model yielding zero scattered intensity, the fit algorithm wil find a compromise characterized by a large damping rate $\Gamma$. Moreover, since the fit cannot probe close to the forbidden boundary of $\omega_{\vec{q}}$= 0, the algorithm will return an uncertaintly in the fit parameters that severely underestimates the real uncertainties. As such, when dealing with broad signals, one should never employ eqn \ref{ah} since this equation is, strictly speaking, only valid for $z _{\vec{q}}< 2f_{\vec{q}}$, and in practice underestimates the errorbars of the fit parameters when $\Gamma_{\vec{q}} \approx \omega_{\vec{q}}$. Bluntly speaking, when one models overdamped excitations using eqn \ref{ah}, then one puts a pendulum under water and equates the time it takes the pendulum to fall back to its equilibrium position with the oscillation time, even though oscillatory motion is no longer supported.\\

\begin{figure}[t]
\begin{center}  
\includegraphics*[viewport=20 100 550 730,width=85mm,clip]{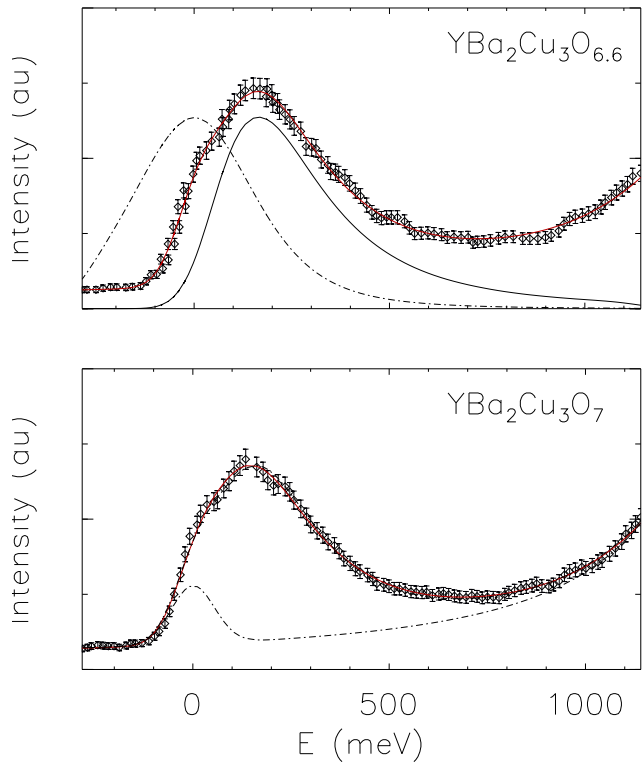}
\end{center}
\caption{(Color online) RIXS spectra\cite{ybco} (points plus errorbars) for YBa$_2$Cu$_{3}$O$_{6.6}$ and YBa$_2$Cu$_{3}$O$_{7}$ measured at $\vec{q}$= (0.13,0,0). The solid line through the data points in both panels is the result of the fitting procedure described in the text that fits a harmonic oscilaltor as well as a background contribution (dashed-dotted curve in the bottom panel) to the data. For both panels, the best fit (lighter curve) corresponds to an overdamped harmonic oscillator. The harmonic contribution has been shown separately in the top panel as a solid curve. We also show this same contribution, but now with the system-independent frequency factor $\omega/(1-e^{-\beta \hbar\omega})$ taken out (dashed-dotted curve centered around E = 0). The dashed-dotted curve through the data points  in the bottom panel is the result of a fit according to eqn \ref{ah} with $\omega_{\vec{q}}$= 250 meV. The two fits are virtually indistinguishable from each other; the quality of this fit is only slighty worse (30\%) than the best fit representing an overdamped harmonic oscillator. Only very close scrutiny reveals that the fit to eqn \ref{ah} slightly underestimates the data in some regions, and overestimates them in others. The comparison between the two fits indicates that even if the paramagnons are propagating, then the errorbars on the propagation frequency must be much larger than reported in reference [\onlinecite{ybco}].} \label{rep}
\end{figure}

It is easy to see why the scattering data can fool one into believing that eqn \ref{ah} would be an appropriate model to use: the paramagnon scattering peaks well away from  $\omega_{\vec{q}}$= 0, apparently ensuring the validity of eqn \ref{ah}. We show an example of RIXS data in Fig. \ref{rep}. However, the peak position at finite energies can be entirly due to the temperature-dependent frequency factor in eqn \ref{db}, which pushes the scattered intensity out to higher energies. We illustrate this by combining eqns \ref{db} and \ref{ah}, and summing the two contributions at $\pm  \omega_{\vec{q}}$ to give (using $\chi_{\vec{q}}$ to indicate the strength of the resonance in the susceptibility):
\begin{equation}
S(\vec{q},\omega)=  \cfrac{\omega\chi_{\vec{q}}}{1-e^{-\beta \hbar\omega}}
\cfrac{4\Gamma_{\vec{q}}\omega_{\vec{q}}} {(\omega^2-\omega_{\vec{q}}^2-\Gamma_{\vec{q}}^2)^2+(2\omega\Gamma_{\vec{q}})^2}.
\label{dhoeqn}
\end{equation}
For low temperatures ($\beta \hbar \omega \gg 1$), the denominator equals 1, and the frequency factor $\omega$ in the numerator  pushes the resonance at $\omega = \omega_{\vec{q}}$ out to higher energies. In the case of overdamped modes, we can no longer use this equation\cite{wouterbook}, but instead we have to use the more general equation:
\begin{equation}
S(\vec{q},\omega)=  \cfrac{\omega\chi_{\vec{q}}}{1-e^{-\beta \hbar\omega}}
\cfrac{2z_{\vec{q}}f_{\vec{q}}} {(\omega^2-f_{\vec{q}} ^2)^2+(\omega z_{\vec{q}})^2}.
\label{dhogen}
\end{equation}
This equation, which has a slightly different definition for $\chi_{\vec{q}}$, is valid for all cases: damped, critically damped and overdamped. Depending on whether $f_{\vec{q}} > z_{\vec{q}}/2$ or vice versa, we find damped or overdamped modes, respectively. The poles of the dynamic susceptibility are located at $i\omega =\pm i\sqrt{f_{\vec{q}}^2-z_{\vec{q}}^2/4}- z_{\vec{q}}/2 $. Similar to eqn \ref{dhoeqn}, the scattered intensity gets pushed out to higher frequencies. Even for the case of overdamped excitations, this results in a peak in $S(\vec{q},\omega)$ at finite energies, and one might be inclined to assume that the system is actually supporting well-defined propagating excitations and resort to eqn \ref{ah} for model fitting. Next, we re-analyze some spectra from the literature\cite{ybco,lasrcuo} and show how the usage of eqn \ref{ah} has led to erroneous results.\\

In a recent paper\cite{ybco}, RIXS was performed on an extended family of high-temperature superconductors. The conclusion inferred from this study was that the entire family of systems supported damped spin wave excitations (paramagnons), with dispersions very similar to magnons in undoped, magnetically ordered cuprates. The authors used this finding for quantitative tests of magnetic Cooper pairing mechanisms. We reproduce two of their spectra in Fig. \ref{rep}. Based on a fit using the model described by eqn \ref{ah}, the authors concluded that both these systems at this particular wave vector supported propagating paramagnons with excitation energy of roughly 125 meV and damping rate (Half Width at Half Maximum-HWHM) of roughly 225 meV. Given the small ratio of the excitation energy compared to the damping rate, this is suggestive of having used eqn \ref{ah} outside of its range of validity.\\

We have re-analysed the data in Fig. \ref{rep} using eqn \ref{dhogen} and find that the paramagnons are overdamped. To do so, we fit the data to a model that has 6 free parameters. Three of the parameters are for $\chi_{\vec{q}}$, $f_{\vec{q}}$, and $  z_{\vec{q}}$, describing the paramagnon, and three parameters decribe the elastic and background scattering. The elastic scattering was modeled by a gaussian located at $E=\hbar\omega$ = 0 with HWHM of 65 mev (as reported in reference [\onlinecite{ybco}]), leaving only its amplitude as a free parameter; the sloping background was modeled by $a +b/(1700-\hbar\omega)^2$, capturing the time independent background and the tails of a peak in the scattered intensity located at $\omega$ = 1700 meV\cite{ybco}. We report the results of the fit in Fig. \ref{rep} and notice that the agreement between model and data is (at least) as good as the agreement reported in reference [\onlinecite{ybco}]. The main difference between the fit published in the literature and the one shown in Fig. \ref{rep} is that the paramagnons are overdamped.\\

Note that the normally straightforward way of removing the frequency prefactor\cite{rot} that pushes the poles of  $\chi(\vec{q},\omega)$  out to higher frequencies does not work in the case of RIXS where the background contribution has to be modeled and subtracted first. In liquids, one normally\cite{wouterbook} analyzes and plots the function $(1-e^{-\beta \hbar\omega}) S(\vec{q},\omega)/\omega$ (which is the Fourier transform of the relaxation function) since this function gets rid of the obfuscating frequency factor, rendering a function that has resonances located at the poles of the dynamic suscpetibility. We show this function for the harmonic oscillator contribution obtained from the model fit in Fig. \ref{rep}. Inspection of this function immediately reveals that the excitations are non-propagating.\\

We stress that it may not really be that important for a system whether excitations are overdamped, or whether they propagate for about one wavelength before they dampen out. However, the {\it changes} in the propagation as a function of sample composition probably are inportant, and these changes can only be teased out of the data in a reliable manner when the data are fitted to a proper model that allows for these changes to happen in the first place.\\ 

We also re-analyze some recent experiments\cite{lasrcuo}  on La$_{2-x}$Sr$_x$CuO$_4$, where RIXS was performed in order to assess the magnetic excitations as a function of doping $x$. Similar to the aforementioned study, the main conclusion of this study was that propgating paramagnons can be found for all concentrations $x$, with excitations energies more or less independent of the doping concentration. The main difference between the various doping levels was found\cite{lasrcuo}  to be the damping rate of the paramagnons. The authors concluded that these findings indicated that the mechanism behind high-temperature superconductivity could not be related to high-energy magnetic excitations. Upon re-analyzing their results, we find that this conclusion might well be premature.\\

\begin{figure}[t]
\begin{center}  
\includegraphics*[viewport=30 130 570 550,width=85mm,clip]{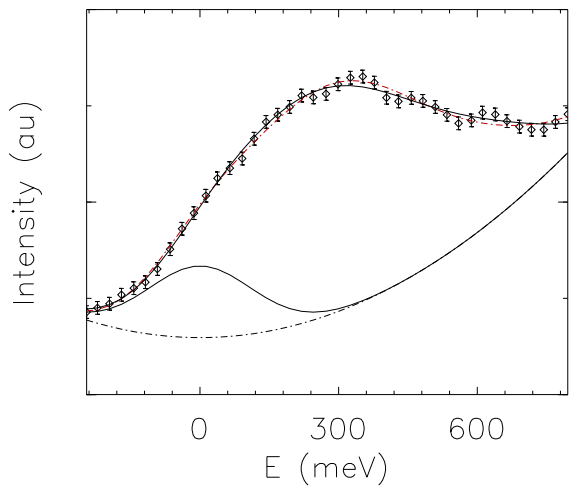}
\end{center}
\caption{(Color online) RIXS spectra\cite{lasrcuo} (points plus errorbars) for La$_{1.6}$Sr$_{0.4}$CuO$_4$ measured at $\vec{q}$= (0.33,0,0). The dashed-dotted curve through the points is the result of a fit to a damped harmonic oscillator with $\omega_{\vec{q}}$= 250 meV and the background given by the solid line below the data points. The solid line through the data points corresponds to an overdamped harmonic oscillator. $\chi^2$ values for the latter are actually better than for the former. Scrutiny of the two fit results shows that the overdamped oscillator (solid curve) captures the behavior of most points slightly better, but the damped oscillator (dashed-dotted curve) does a slightly better job around the peak position. What is clear is that while either scenario describes the data well, the background and accuracy of the points dictates which scenario is favored by the fitting routine.} \label{rep2}
\end{figure}

The results of our re-analysis of the published spectra\cite{lasrcuo} indicate that paramagnons might well be propagating for all $q$-values and concentrations, but that the errorbars on the propagation frequencies are much larger than those publsihed, and that it actually is possible that some paramagnons are overdamped (see Fig. \ref{rep2}). This RIXS study had a worse energy resolution than the one discussed previously (HWHM of 140 meV versus 65 meV), and the background appeared to be more difficult to model. As such, we find that the fit parameters depend strongly on the exact model used for the background. In the published analysis\cite{lasrcuo}, the background played only a minor role in identifying $\omega_{\vec{q}}$ since the fit was prohibited from straying too close to $\omega_{\vec{q}}$=0. Using eqn \ref{dhogen} instead, we find that the difference between propagating excitations and overdamped ones is determined by the flanks of the resonance. As a consequence, the exact modeling of the background now plays a major role. We show the results in Fig. \ref{rep2}.\\

The moral for the case study described above is, that in order to extract detailed information about the paramagnons, one needs to do a better job in modeling the background. For instance, allowing for different backgrounds in all 25 spectra published in reference [\onlinecite{lasrcuo}], while probing the system with an energy resolution comparable to the widths of the excitations one is interested in, does not allow for firm conclusions to be drawn when it comes to details of doping dependence (Fig. \ref{rep2}). While the usage of eqn \ref{ah} might yield well-defined fit parameters, in reality the degree of uncertaintly is much larger (see Fig. \ref{rep2}). As pointed out, the true degree of uncertainty can only be assessed when the fit is allowed to probe the region forbidden by eqn \ref{ah}. In our opinion, the resulting errorbars on the paramagnon excitation energies do not allow for the firm conclusions drawn in reference [\onlinecite{lasrcuo}]; the data, more than likely, show a significant change from well-defined propagating modes in the anti-ferromagnetic compound, to strongly damped or overdamped modes for different concentrations $x$. We find that, at best, the data are consistent with the conclusions drawn by the authors\cite{lasrcuo}, but do not lead to them. The study probably needs to be re-analyzed using the proper model, and preferably, by employing a higher energy resolution.\\

As a final word of caution, fitting the spectra to a sum of Gaussian lineshapes as is sometimes done in the analysis of RIXS data (see, for example, reference [\onlinecite{gaussian})] suffers from the same drawback as fitting the data to eqn \ref{ah}: it is assumed a priori that the peak position in the spectra corresponds to an actual pole of the dynamic suscpetibility rather than to a feature pushed out to higher energy transfers because of the action of a temperature-dependent prefactor linking the dynamics susceptibility to the scattered intensity.\\

 In conclusion, we have shown that the usage of a model outside its range of validity has led to conclusions that are not (fully) justified by the data. Given the amount of effort\cite{rixs} that has gone into establishing RIXS as a new and complementary technique for investiagating paramagnons, we strongly suggest to no longer use a model that assumes a particular outcome beforehand, but rather to use the general model for the harmonic oscillator that allows for excitations to be overdamped.


\begin{thebibliography}{10}
\bibitem{rixs} Luuk J. P. Ament, Michel van Veenendaal, Thomas P. Devereaux, John P. Hill, and Jeroen van den Brink, Rev. Mod. Phys. {\bf 83}, 705 (2011).
\bibitem{wouterbook} For example, see Chapter 3 of Wouter Montfrooij and Ignatz de Schepper, {\it Excitations in simplle liquids, liquid metals and superfluids} (Oxford university Press, Oxford, 2010).
\bibitem{rot} E.C. Svensson, W. Montfrooij, and I.M. de Schepper, Phys. Rev. Lett.  {\bf 77},
4398 (1996).
\bibitem{disclaimer} The references below merely represent a selection of recent articles where the existence of propagating modes is assumed beforehand. This selection is neither targeted at any group in particular, nor is it exhaustive.
\bibitem{ybco} M. Le Tacon, G. Ghiringhelli, J. Chaloupka, M. Moretti Sala, V. Hinkov, M.W. Haverkort,
M. Minola, M. Bakr, K. J. Zhou, S. Blanco-Canosa, C. Monney, Y. T. Song, G. L. Sun, C. T. Lin,
G. M. De Luca, M. Salluzzo, G. Khaliullin, T. Schmitt, L. Braicovich, and B. Keimer, Nature Physics   {\bf 7}, 725–730 (2011).
\bibitem{lasrcuo}M. P. M. Dean, G. Dellea, R. S. Springell, F. Yakhou-Harris, K. Kummer, N. B. Brookes, X. Liu,
Y-J. Sun, J. Strle, T. Schmitt, L. Braicovich, G. Ghiringhelli, I. Bozovi\'{c}, and J. P. Hill, Nature Materials   {\bf 12}, 1019–1023 (2013).
\bibitem{cupa} M. Le Tacon, M. Minola, D. C. Peets, M. Moretti Sala, S. Blanco-Canosa, V. Hinkov, R. Liang, D. A. Bonn, W. N. Hardy, C. T. Lin, T. Schmitt, L. Braicovich, G. Ghiringhelli, and B. Keimer, Phys. Rev. B   {\bf 88}, 020501 (R) (2013).
\bibitem{pnictides} Ke-Jin Zhou,	Yao-Bo Huang,	Claude Monney,	Xi Dai,	Vladimir N. Strocov,	Nan-Lin Wang,	Zhi-Guo Chen,	Chenglin Zhang,	Pengcheng Dai,	Luc Patthey,	Jeroen van den Brink,	Hong Ding, and Thorsten Schmitt, Nature Communications   {\bf 4}, 1470 (2013).
\bibitem{bis} M. P. M. Dean, A. J. A. James, R. S. Springell, X. Liu, C. Monney, K. J. Zhou, R. M. Konik, J. S. Wen, Z. J. Xu, G. D. Gu, V. N. Strocov, T. Schmitt, and J. P. Hill, Phys. Rev. Lett.   {\bf 110}, 147001 (2013).
\bibitem{bis2}Y. Y. Peng, M. Hashimoto, M. Moretti Sala, A. Amorese, N. B. Brookes, G. Dellea, W.-S. Lee, M. Minola, T. Schmitt, Y. Yoshida, K.-J. Zhou, H. Eisaki, T. P. Devereaux, Z.-X. Shen, L. Braicovich, and G. Ghiringhelli, Phys. Rev. B   {\bf 92}, 064517 (2015).
\bibitem{gaussian} S. Wakimoto, K. Ishii, H. Kimura, M. Fujita, G. Dellea, K. Kummer, L. Braicovich, G. Ghiringhelli, L. M. Debeer-Schmitt, and G. E. Granroth, Phys. Rev. B   {\bf 91}, 184513 (2015).
\end{thebibliography}
\end{document}